\documentclass{article}                    
\usepackage{authblk}
\usepackage{algorithm,algpseudocode}
\usepackage{url}
\usepackage{graphicx}
\usepackage{amssymb}
\usepackage{amsmath}
\usepackage[para,online,flushleft]{threeparttable}
\usepackage{multirow}
\usepackage{caption}
\usepackage{subcaption}
\usepackage{booktabs}
\usepackage[numbers]{natbib}
\usepackage{csquotes}
\usepackage{enumitem}

\graphicspath{ {images/} }

\begin{document}

\title{Optimization of Heterogeneous Systems with AI Planning Heuristics and Machine Learning: A Performance and Energy Aware Approach}

\author[$1$]{Suejb Memeti}
\author[$2$]{Sabri Pllana}

\affil[$1$]{Blekinge Tekniska Hogskola, Department of Computer Science, 371 79 Karlskrona, Sweden; E-mail: suejb.memeti@bth.se}   
\affil[$2$]{Linnaeus University, Department of Computer Science and Media Tech., 351 95 Vaxjo, Sweden; E-mail: sabri.pllana@lnu.se}   

\date{Preprint}  

\maketitle

%===========================================================================

\begin{abstract}
Heterogeneous computing systems provide high performance and energy efficiency. However, to optimally utilize such systems, solutions that distribute the work across host CPUs and accelerating devices are needed. 
In this paper, we present a performance and energy aware approach that combines AI planning heuristics for parameter space exploration with a machine learning model for performance and energy evaluation to determine a near-optimal system configuration. For data-parallel applications our approach determines a near-optimal host-device distribution of work, number of processing units required and the corresponding scheduling strategy. We evaluate our approach for various heterogeneous systems accelerated with GPU or the Intel Xeon Phi. The experimental results demonstrate that our approach finds a near-optimal system configuration by evaluating only about 7\% of reasonable configurations. Furthermore, the performance per Joule estimation of system configurations using our machine learning model is more than 1000x faster compared to the system evaluation by program execution.  

\end{abstract}

%=====================================================================
\section{Introduction}
\label{sec:introduction}

Accelerators (such as, GPU or Intel Xeon Phi) are often used collaboratively with general-purpose CPUs to increase the overall system performance and improve energy efficiency. Currently, three of the top five most powerful computers in the TOP500 list \cite{top500} are heterogeneous computers that use GPUs as accelerators. Efficient programming of heterogeneous systems and splitting the work between the CPUs and accelerators are domains of particular interest \cite{Czarnul:2018,Markidis:2018,Amaral:2020}.  

Due to different architectural characteristics and the large number of system parameter configurations (such as, the number of threads, thread affinity, workload partitioning between multi-core processors of the host and the accelerating devices), achieving a good workload distribution that results with optimal performance and energy efficiency on heterogeneous systems is a non-trivial task \cite{vettersurvey,pllana:2017}. An optimal system configuration that results with the highest throughput may not necessarily be the most energy efficient. Furthermore, the optimal system configuration is most likely to change for different types of applications, input problem sizes, and available resources. 

\begin{figure*}
     \centering
     \begin{subfigure}[t]{0.45\textwidth}
         \centering
         \includegraphics[width=5cm]{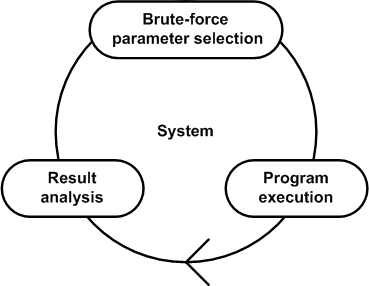}
          \caption{Brute-force search}
         \label{fig:brute}
     \end{subfigure}%
     ~ 
     \begin{subfigure}[t]{0.45\textwidth}
         \centering
         \includegraphics[width=5cm]{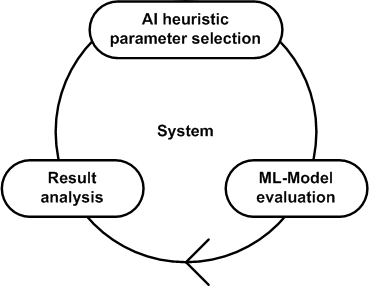}
         \caption{AI heuristic search}
         \label{fig:heuristic}
     \end{subfigure}
        \caption{System performance tuning. (a) Evaluation of all possible configurations with program execution, (b) AI heuristic selection of candidate configurations and model-based evaluation.}
        \label{fig:tuning}
\end{figure*}

Figure \ref{fig:tuning} depicts the process of system performance tuning. Traditionally the process of finding the optimal system parameters involved many iterations of selecting parameter values, program execution, and performance analysis (Figure \ref{fig:brute}). A brute-force search requires the program execution for all parameter values of interest, and consequently for real-world programs and systems it may take an unreasonable long time to find the optimum. In contrast to brute-force search, an AI heuristic search \cite{Alba:2018,NumRecipes2007} enables to find a near-optimum solution with fewer performance experiments based on nature-inspired algorithms (for instance, simulated annealing \cite{Kirkpatrick:1983} or Artificial Bee Colony \cite{Karaboga:2007}). Figure \ref{fig:heuristic} depicts our approach that combines AI heuristic parameter value selection with a machine learning model for performance evaluation. The advantage of our approach is two-fold: 

\begin{itemize}
    \item heuristic parameter value selection reduces significantly the number of performance experiments compared to brute-force search;
    \item system performance evaluation using a model is usually much faster than the program execution on a real-world system; moreover, availability of the system under study during the performance optimization process is not required once that the model has been developed. 
\end{itemize}
 
In \cite{Memeti:2018} authors provide a comprehensive review of literature that describes various approaches for using heuristics or machine learning for optimization of parallel computing systems from single-node systems \cite{grewe2011static,kasichayanula2012power} to large scale Grid and Cloud computing infrastructures \cite{Fahringer:2005,Grzonka:2018,PllanaBMNX08,Xhafa:2009,fahringer04,pbb08}. The related work usually uses for system optimization either heuristics \cite{Pereira:2020} or machine learning \cite{grewe2011static} and there is insufficient research in combining these two methods in particular for optimization of heterogeneous computing systems. Preliminary results of our optimization approach are discussed in \cite{memeti:icppw16}. However, \cite{memeti:icppw16} was limited to only performance optimization of systems accelerated with the Intel Xeon Phi. This paper extends our previous work as follows, 
\begin{itemize}
    \item we present a new machine learning model that estimates the performance per Joule of the system (that is, it considers also energy consumption, beside the performance);
    \item  we demonstrate that our approach is general-applicable and it is not limited to Intel Xeon Phi, by applying our method also to GPU accelerated systems; 
    \item we demonstrate the applicability of our approach with two applications: Pearson correlation coefficient and parallel pattern matching. 
\end{itemize}

We use an AI heuristic search technique for parameter space exploration. The optimization process involves generation of system configuration based on random selection of parameter values and the system performance evaluation using a machine learning model. Our method guides the process of intelligent navigation through the parameter space towards the determination of the near-optimal system configuration. As a result, our method requires only a small fraction of possible performance experiments. The empirical evaluation of our approach demonstrates that by using only about 7\% of the total brute-force experiments we are able to determine a near-optimal system configuration with respect to the performance per Joule. Moreover, using our machine learning model for evaluation of system configurations is more than 1000x faster than the system evaluation by program execution.

This paper is organized as follows. Section \ref{sec:sysapps} introduces computing systems and applications that we use to illustrate our approach. We describe our optimization approach in Section \ref{sec:approach} and experimentally evaluate it in Section \ref{sec:evaluation}. We discuss the related research in Section \ref{sec:rw}. Section \ref{sec:conclusion} provides a summary of the paper and a description of the future work.  

%===============================================================
\section{Heterogeneous Systems and Applications}
\label{sec:sysapps}

In this section, we describe heterogeneous computing systems and applications that we use in this paper to illustrate and evaluate our approach. In addition, we describe input data-sets and parameters that define system configurations.

%---------------------------------------------------------------
\subsection{Heterogeneous Computing Systems}
\label{sec:sys}

Table \ref{table:emil_and_ida} lists the major properties of heterogeneous computing systems \textit{Emil} and \textit{Ida} that we use for experimentation in this paper. 

\begin{table}[ht]
	\scriptsize
	\caption{Major features of Ida and Emil.}
	\label{table:emil_and_ida}
	\centering
	\begin{tabular}{lll|ll}
  \hline
   & \multicolumn{2}{c|} {Ida} & \multicolumn{2}{c}{Emil} \\ 
   & Intel Xeon E5 & GeForce GPU & Intel Xeon E5 & Intel Xeon Phi \\
  \hline
	Processor & E5-2650 v4 & GTX Titan X & E5-2695 v2 & 7120P \\
	Core Frequency & 2.2 - 2.9 GHz &  1 - 1.1 GHz &  2.4 - 3.2 GHz &  1.2 - 1.3 GHz \\
	Number of cores & 12 & 3072 & 12 & 61 \\
	Number of threads & 24 & / & 24 & 244 \\
	Cache size &  30 MB &  / &  30 MB &  30.5 MB \\
	Mem. Bandwidth &  76.8 GB/s & 336.5 GB/s & 59.7 GB/s & 352 GB/s \\
	Memory size &  384 GB & 12 GB & 128 GB & 16 GB \\
	TDP &  105 W & 250 W & 115 W & 300 W \\			
  \hline
	\end{tabular}
\end{table}

\textit{Ida} comprises two Intel Xeon E5-2650 v4 general purpose CPUs on the host, and one GeForce GTX Titan X GPU as accelerator. Each CPU has 12 physical cores, and each core of CPU supports two threads. In total, the host CPUs provide 24 cores and 48 threads. The GPU device has 24 Streaming Multiprocessors (SM), and in total 3072 CUDA cores running at base frequency of 1 GHz. 

\textit{Emil} comprises two Intel Xeon E5-2695 v2 general purpose CPUs on the host, and one Intel Xeon Phi 7120P co-processing device. Similar to \textit{Ida}, Each CPU of \textit{Emil} has 12 cores that support two threads per core (known as logical cores) that amounts to a total of 48 threads. The Xeon Phi device has 61 cores running at 1.2 GHz base frequency; each core supports four hardware threads. In total, the Xeon Phi supports 244 threads. The lightweight Linux operating system installed on the Xeon Phi device runs on one of the cores. 

The parameters that define the system configuration for our optimization approach are shown in Table \ref{table:parameter-values}. All the parameters are discrete. 

The considered system parameters for the PCC application executed on \textit{Ida} include the CPU workload and GPU workload. The workload fraction parameter can have any number in the range $\{0,..,100\}$, such that if $60\%$ of the workload is assigned for processing to the host CPU, the remaining $100 - 60 = 40\%$ is assigned to the GPU device. 

The considered parameters for the pattern matching application executed on \textit{Emil} include the number of CPU threads and accelerator threads, the CPU and accelerator thread affinity, and the CPU and accelerator workload fraction. The system parameter values for the host CPU threads are $\{12, 24, 36, 48\}$, whereas for the accelerator are $\{60, 120, 180, 240\}$. The thread affinity can vary between \{none (0), compact (1), scatter (2)\} for on host CPUs, and \{balanced (0), compact (1), scatter (2)\} on accelerator. Similar to the Ida system, the CPU and accelerator workload fraction can have values in the range $\{0,..,100\}$.

\begin{table}[ht]
	\centering
	\scriptsize
	\caption{The set of considered parameters and their values for our target systems.}
	\label{table:parameter-values}
	\begin{tabular}{@{}llll@{}}
		\toprule
		System & Parameters								& Parameter values							\\ \midrule
		
		\multirow{2}{*}{Ida} & CPU workload fraction (CPU-W) &  \{0...100\} \\
		 & GPU workload fraction (GPU-W) & \{100 - CPU-W\} \\ \midrule
		 
		 \multirow{6}{*}{Emil} & CPU threads (CPU-T)     				& \{12, 24, 36, 48\}      			\\
		& Accelerator threads (ACC-T) 			& \{60, 120, 180, 240\}	\\
		& CPU thread affinity (CPU-A)				& \{none (0), scatter (1), compact (2)\} 				\\
		& Accelerator thread affinity (ACC-A) 	& \{balanced (0), scatter (1), compact (2)\} 			\\
		& CPU workload fraction (CPU-W)  			& \{0...100\} 								\\
		& Accelerator workload fraction (ACC-W)	& \{100 - CPU-W\}							\\ \bottomrule
	\end{tabular}
\end{table}

%----------------------------------------------------------------
\subsection{Applications}
\label{sec:apps}

In this section, we describe two applications that we have prepared for execution in hybrid mode (that is, the workload is shared between the host CPUs and the accelerating device): (1) Pearson Correlation Coefficient (PCC) and (2) parallel pattern matching. While the PCC application is adapted to GPU-accelerated systems, the pattern matching application targets systems accelerated with the Intel Xeon Phi. 

%----------------------------------------------------------------
\subsubsection{Pearson Correlation Coefficient (PCC)}
\label{sec:pcc}

The Pearson Correlation Coefficient \cite{boslaugh2012statistics}, which is also known as Pearson product-moment correlation, is a statistical method for measuring the strength of a relationship between two variables \textit{x} and \textit{y}. The correlation may be positive (1), negative (-1), or there is no correlation (0). Positive correlation means that if variable \textit{x} is increased, there will be an increase on the \textit{y} variable, whereas the negative correlation means that an increase on \textit{x} would result with a decrease on \textit{y}. A value of \textit{0} for the correlation means that the two variables are not related and increase on one variable has no effect on the other one. 

The strength of the relationship is measured using the absolute value of the correlation, for instance, \textit{abs(-0.23) = 0.23} is a stronger relationship than \textit{0.2}.  

The Pearson correlation coefficient is calculated using the following formula \cite{eslami2017gpu}: 
\begin{equation}
    \rho_{xy} = \frac{\sum_{i=1}^{n}(x_{i} - \bar{x})(y_{i} - \bar{y})))}{\sqrt{\sum_{i=1}^{n}(x_{i} - \bar{x})^2)}\sqrt{\sum_{i=1}^{n}(y_{i} - \bar{y})^2)}}
\end{equation}

The Pearson coefficient correlation has a wide range of applications. For instance, \cite{eslami2017gpu} show its application to Magnetic Resonance Imaging (MRI) for analyzing the functional connectivity of different brain regions. 

In this paper, we apply the Pearson coefficient correlation to rows of matrices. Note that for each row, we calculate the relationship with all other rows below in the matrix. We added the glue code to the CPU and GPU PCC implementations for enabling the hybrid execution. The input matrix is split row-wise between the CPUs on the host and GPU. 

It is worth to mention, that for this specific application, a \textit{50\% - 50\%} workload distribution does not mean that the work is equally divided between the host and the accelerator. This is because, for instance row \textit{0} compares itself with rows \textit{1, 2, 3 ... n}. However, row \textit{50} compares itself only with rows \textit{51, 52, ... n}. As the program execution progresses through the rows of the matrix, there is less and less work to perform. 

%----------------------------------------------------------------
\subsubsection{Parallel Pattern Matching}
\label{sec:pm}

Pattern matching is the process of finding multiple and overlapping occurrences of sub-strings in a string. Common uses of pattern matching are in search and replace functions, determining the location of a pattern, or highlighting important information out of huge data sets \cite{Vitabile2019}. Pattern matching is also used in intrusion detection systems (to identify potential threats) or computational biology (to identify the location of some patterns in large DNA sequences). 

In this paper, we use a pattern matching implementation \cite{mp-cse14,mp-pbio15} to find sub-sequences of strings in the real-world human DNA sequence (3.17GB) obtained from the GenBank sequence database of the National Center for Biological Information \cite{GenBank}. 

Please note that the \textit{50\%-50\%} workload distribution of the pattern matching application means that the workload is equally divided between the host and the accelerator. 

%===========================================================================
\section{Our Approach for Optimization of Heterogeneous Systems}
\label{sec:approach}

In this section, we first motivate with one experiment the need for systematic optimization approaches, and thereafter we describe our approach for optimization of heterogeneous computing systems that combines AI heuristic search techniques with machine learning. 

% --------------------------------------------------------------------------
\subsection{Motivational Experiment}
\label{sec:motivation}

Heterogeneous computing systems may comprise several CPUs and accelerators. Optimal work-sharing among available CPUs and accelerators is not obvious. Furthermore, considering both execution time and energy consumption makes the work-sharing problem more complex. 

\begin{figure}
     \centering
     \begin{subfigure}[b]{\textwidth}
         \centering
         \includegraphics[width=\textwidth]{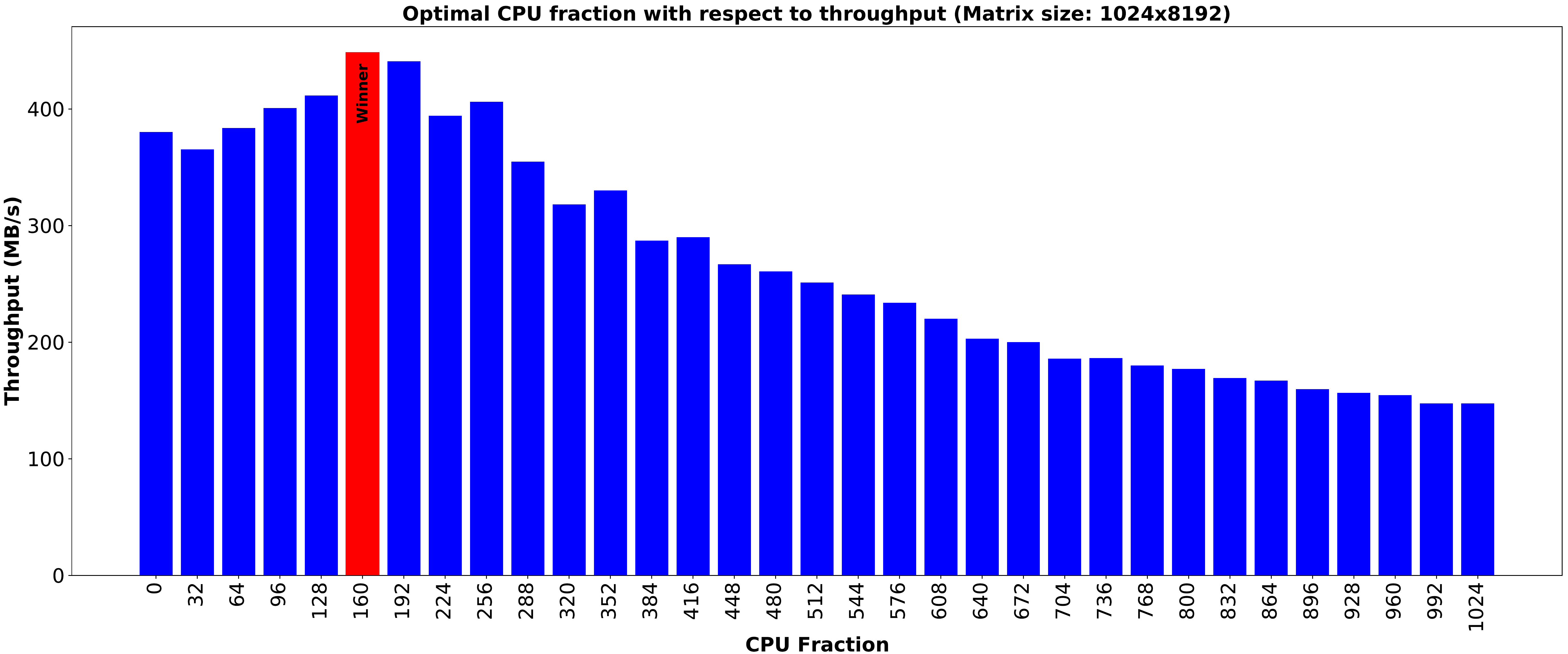}
          \caption{Max throughput is achieved when 160 rows are mapped to CPU and the rest to GPU.}
         \label{fig:mot_throughput}
     \end{subfigure}
     \hfill
     \begin{subfigure}[b]{\textwidth}
         \centering
         \includegraphics[width=\textwidth]{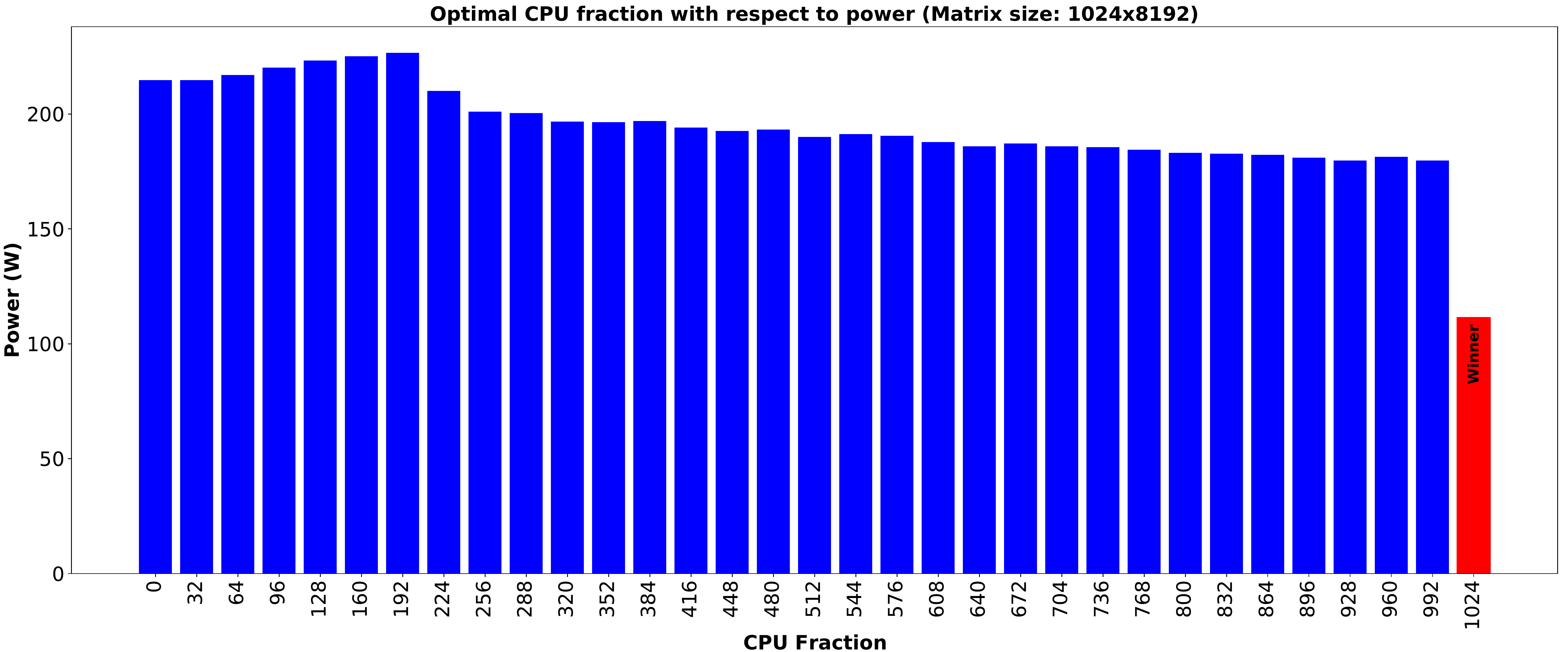}
          \caption{All 1024 rows of matrix mapped to CPU results with the minimal power consumption. Data transfer between CPU and GPU is prohibitive.}
         \label{fig:mot_power}
     \end{subfigure}
     \hfill
     \begin{subfigure}[b]{\textwidth}
         \centering
         \includegraphics[width=\textwidth]{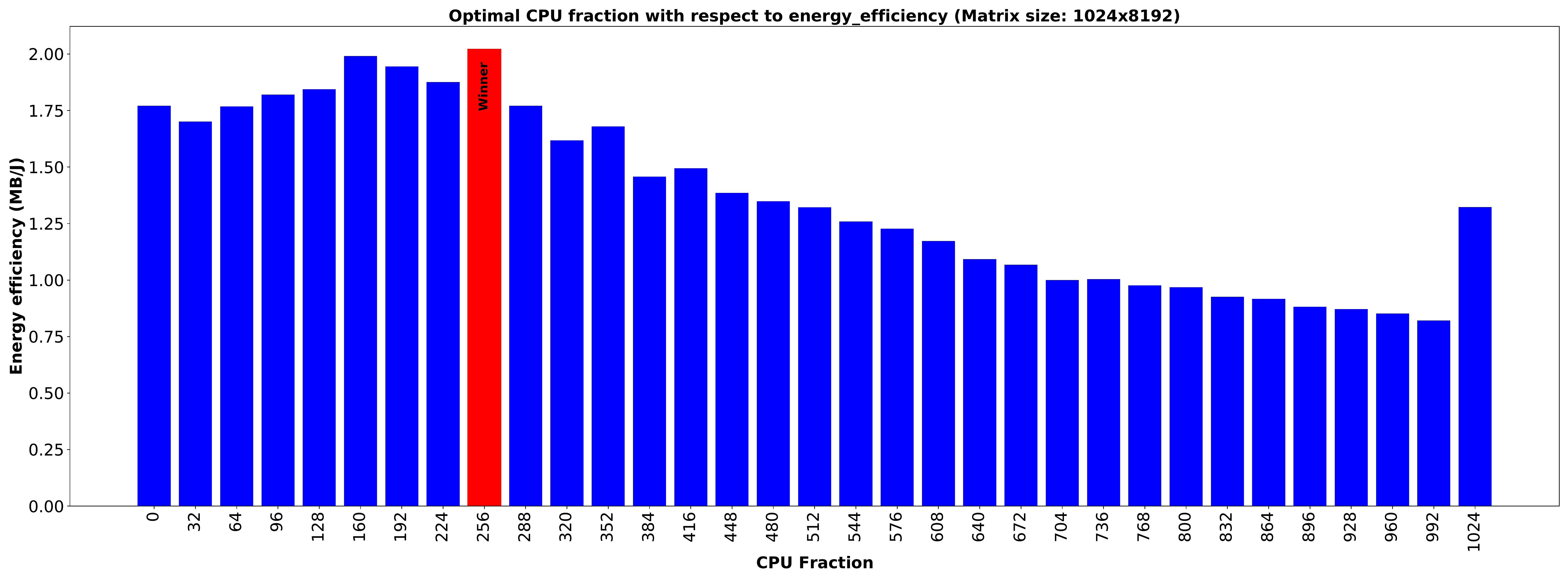}
         \caption{Max energy efficiency is achieved when 256 rows are mapped to CPU and the rest to GPU.}
         \label{fig:mot_power_efficiency}
     \end{subfigure}
        \caption{The CPU fraction that delivers the highest throughput (MB/s), lowest power consumption (W), or highest energy efficiency (MB/J) is not the same. The figure shows the experiments for the PCC application running on Ida for matrix size 1024x8192. Measurements include also the performance overhead for data transfers between CPU and GPU.}
        \label{fig:motivation}
\end{figure}

Figure \ref{fig:motivation} illustrates this problem with the PCC application running on the Ida heterogeneous system that comprises two general-purpose CPUs and one GPU as accelerator. For simplicity, we only show the results for a particular input size (that is a matrix with 1024 rows and 8192 columns) and we only vary the parameter CPU Fraction (that is the number of rows of the matrix mapped to the CPU). Furthermore, we consider only CPU Fraction values that are product of 32. 

From this experiment, we may derive the following insights:
\begin{itemize}
	\item When we consider the $Throughput~[MB/s]$ only, the optimal value for the CPU fraction is 160 rows and rest of the work is mapped to the GPU (Figure \ref{fig:mot_throughput}).
	\item When optimizing for the $Power~[W]$ consumption, the execution on the CPU only is preferable (Figure \ref{fig:mot_power}); please note that also the overhead of data transfers between CPU and GPU is considered for optimization.
	\item When optimizing for $Energy~Efficiency~[MB/J]$ that considers both the throughput and the power consumption, the optimal value for the CPU fraction is 256 (Figure \ref{fig:mot_power_efficiency}).
\end{itemize}

%---------------------------------------------------------------------------
\subsection{System Optimization with AI heuristics and Machine Learning (AML)}
\label{sec:aml}

In this section, we propose an intelligent approach for determination of near-optimal system configuration using AI heuristic search techniques and machine learning. We contrast traditional approaches (also known as brute-force search) that we refer to as Enumeration and Measurements (EM), with our approach of combining AI heuristics with Machine Learning (AML).

\begin{figure}[ht]
	\centering
		\includegraphics[width=\textwidth]{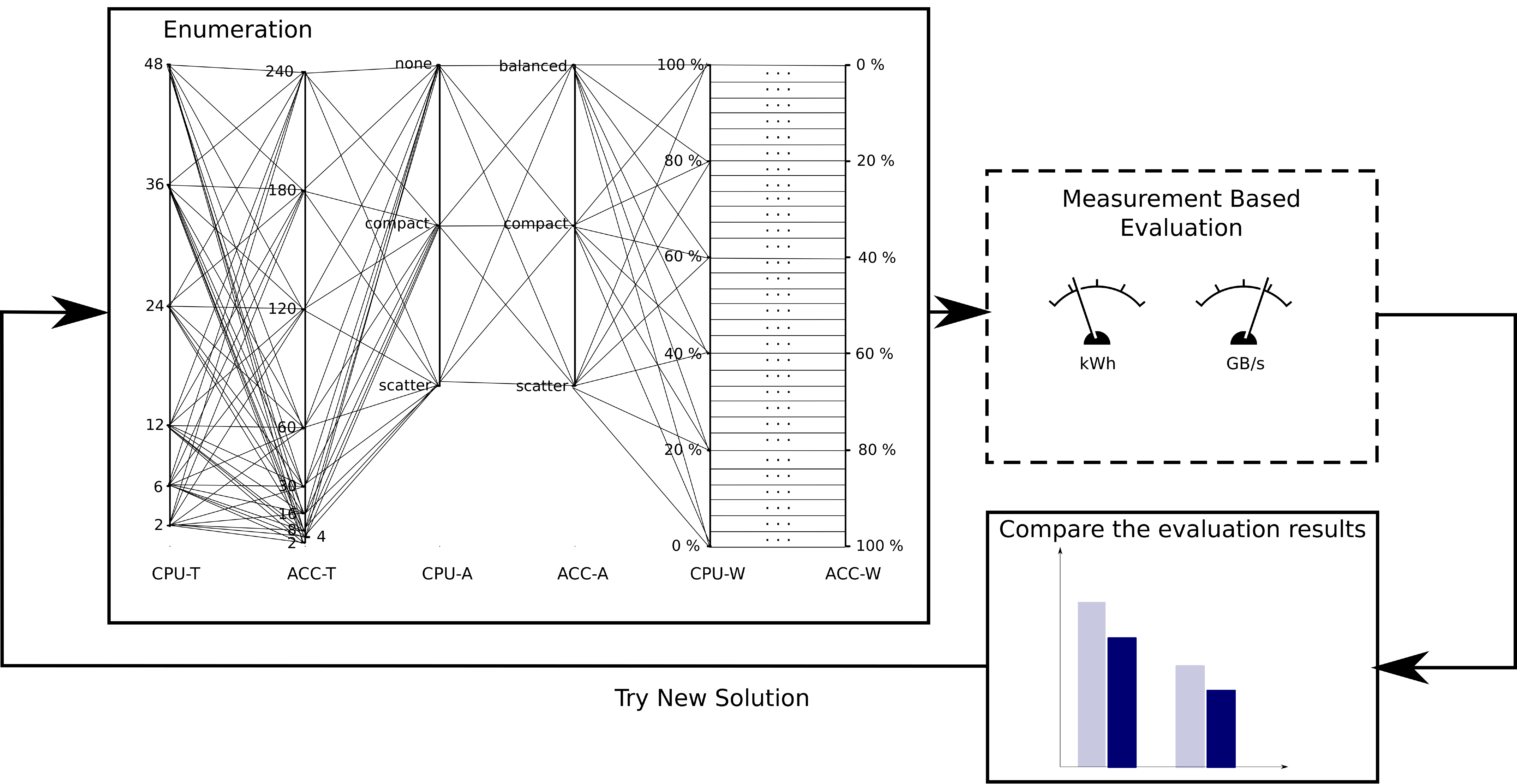}
	\caption{Enumeration and measurements may help to determine the optimal system configuration, however the required effort may be high due to the large parameter space.}
	\label{fig:em-process}
\end{figure}

The EM technique (also known, as brute-force search) is depicted in Figure \ref{fig:em-process}. Compartment \emph{Enumeration} visualizes the process of selecting each feasible value of various system parameters. Thereafter, for each combination of parameter values the system performance is measured. Basically, EM involves program execution on a real computing system for all feasible system parameter values. For real-world programs and systems, EM may result with a very large number of performance measurement experiments. Furthermore, program execution on real computing systems may be time-consuming. Usually computing systems are shared among various users and it may be impractical to block other users for an unreasonable large amount of time while we execute our performance experiments. Key EM drawbacks include,

\begin{itemize}
    \item a prohibitively large number of performance experiments,
    \item requires a dedicated access to the system under study.
\end{itemize}

Determining the optimal system configuration using brute-force may be prohibitively time expensive. The number of all possible system configurations is a product of parameter value ranges, 

\begin{equation} \label{eq:1}
\prod_{i=1}^{n}	R_{c_i} = R_{c_1} \times R_{c_2} \times .. \times R_{c_n}
\end{equation}

where $C=\{c_1, c_2, ..., c_n\}$ is a set of parameters and each $c_i$ has a value range $R_{c_i}$.

\begin{figure}[ht]
	\centering
	\includegraphics[width=\textwidth]{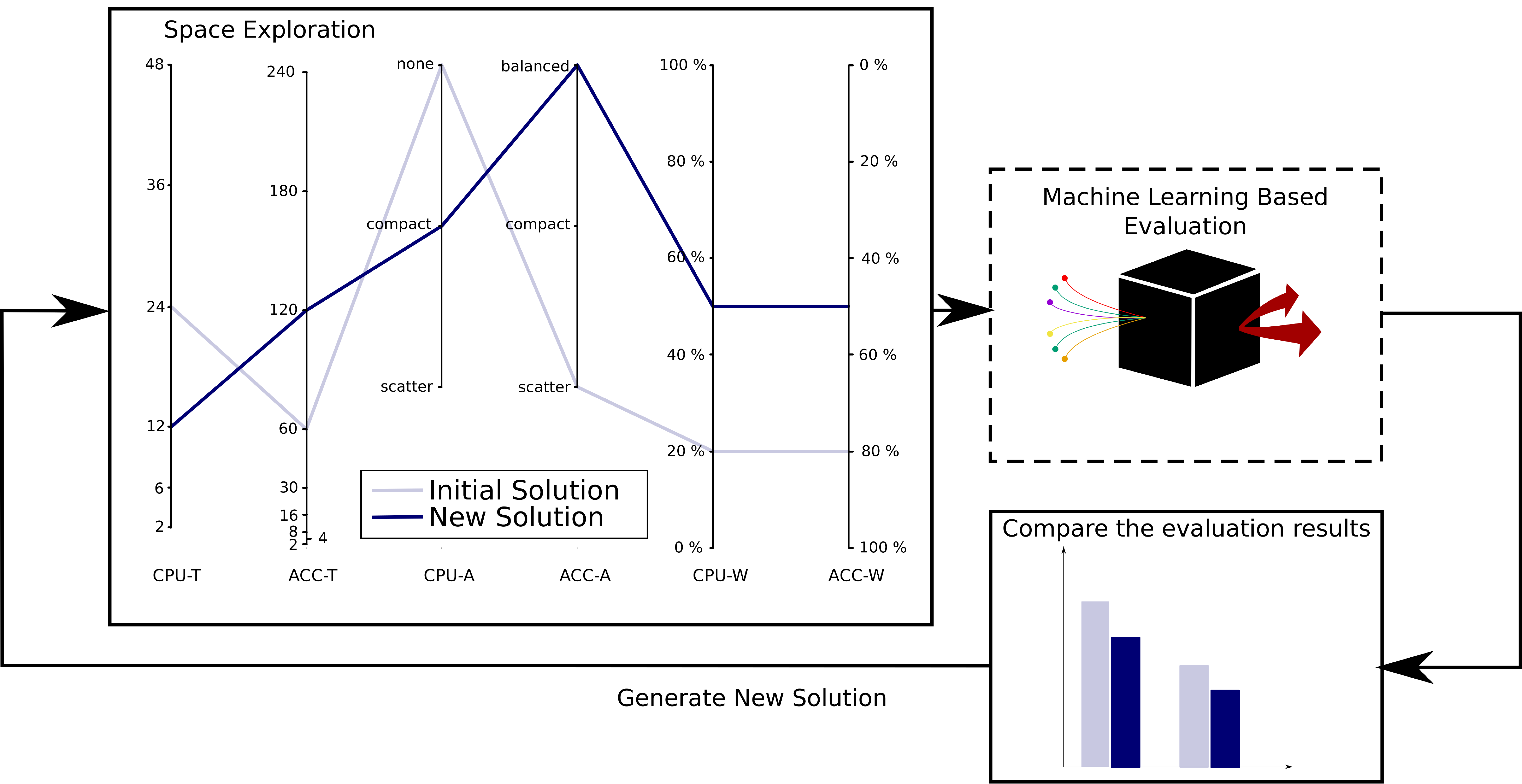}
	\caption{Our approach uses an AI heuristic search technique for design space exploration and machine learning for system performance evaluation.}
	\label{fig:our-approach}
\end{figure}

The aim is to find an optimization approach that is able to find a near-optimal system configuration without having to measure the system performance for all feasible parameter values. Our approach AML is depicted in Figure \ref{fig:our-approach}. Key AML features include,

\begin{itemize}
    \item requires only a small fraction of possible performance experiments,
    \item uses a machine learning model for performance evaluation.
\end{itemize}

AML uses a AI heuristic search technique for parameter space exploration. The optimization process involves generation of system configuration based on random selection of parameter values and the system performance evaluation using a machine learning model. AML guides the process of intelligent navigation through the parameter space towards the determination of the near-optimal system configuration. As a result, AML requires only a small fraction of possible performance experiments. 

In what follows, we describe into more details the two most important components of our approach, which are the Simulated Annealing heuristic for design space exploration and boosted decision tree machine learning algorithm for system performance evaluation. 

% ------------------------------------------------------------------
\subsubsection{Using Simulated Annealing (SA) for Parameter Space Exploration}
\label{sec:SA}

Simulated Annealing (SA) \cite{Kirkpatrick:1983} algorithm is a probabilistic optimization technique for approximately determining the global optimum of a function. SA is a popular technique for searching for optimum in a discrete space, where gradient-based methods are not applicable. An important feature of SA is the capability to find a global optimum also for functions that have many local optimums. 

SA is inspired from the process of annealing in metallurgy, a technique that requires heating and controlled cooling of materials \cite{NumRecipes2007}. At high temperatures $T$ particles of the material have more freedom of movement, and as the temperature decreases the movement of particles is restricted. When the material is cooled slowly, the particles are ordered in the form of a crystal that represents minimal energy state of the material.

\begin{algorithm}
\caption{Our SA-based algorithm for exploration of system configuration space.}
\label{alg:sa}
\begin{algorithmic}[1]
\State {$T$}: SA parameter for accepting a system configuration as improvement
\State {$x$}: intermediate system configuration  
\State {$winner$}: optimal system configuration
\State Initially evaluate CPU only and accelerator only solutions
\State Set the initial temperature {$T$}
\State Randomly generate the initial system configuration {$x$}
	\While{$T > 1$}
		\State Generate a new solution {$x$'} by randomly modifying {$x$}
		\State Evaluate {$x$'} using our machine learning model
		\If{$x$' better than $x$ or acceptance criterion is met}
			\State Replace {$x$} with {$x$'}
			\State update {$winner$}
		\EndIf
		\State Decrease {$T$}
	\EndWhile
\end{algorithmic}
\end{algorithm}

Algorithm \ref{alg:sa} shows the pseudo-code of SA-based method for system configuration space exploration. A system configuration is determined by specific parameter values (see Table \ref{table:parameter-values}.) Since the optimal workload distribution for some of the experiments is to either run everything on the CPU or GPU, we initially evaluate the CPU and accelerator only system configurations (line 4). The process starts by initializing the temperature $T$ (line 5) and generating a random system configuration $x$ (line 6). New solutions $x'$ are generated by randomly modifying the parameter values of current solution $x$ (line 8). We evaluate each of the generated system configurations using our machine learning based prediction model (line 9). If the performance per Joule of $x'$ is better than the one of $x$, we replace $x$ with $x'$ unconditionally (line 11), otherwise we consider accepting $x'$ based on probability $p$ that is described by Equation \ref{eq:botlzmann}. 

\begin{equation} \label{eq:botlzmann}
p = exp((x - x') / T)
\end{equation}

The acceptance criterion (also known as Boltzmann's probability distribution \cite{NumRecipes2007}) allows simulated annealing to get out of local optimums in favor of a global optimum. The temperature variable plays a decisive role in the acceptance criterion. If the temperature $T$ is high, the system is more likely to accept solutions that are worse than the current one. After each new solution we decrease the temperature (line 14). Steps 8-14 are repeated as long as the temperature $T$ is greater than one (line 15).

\begin{figure}[!ht]
	\centering
	\includegraphics[width=\linewidth]{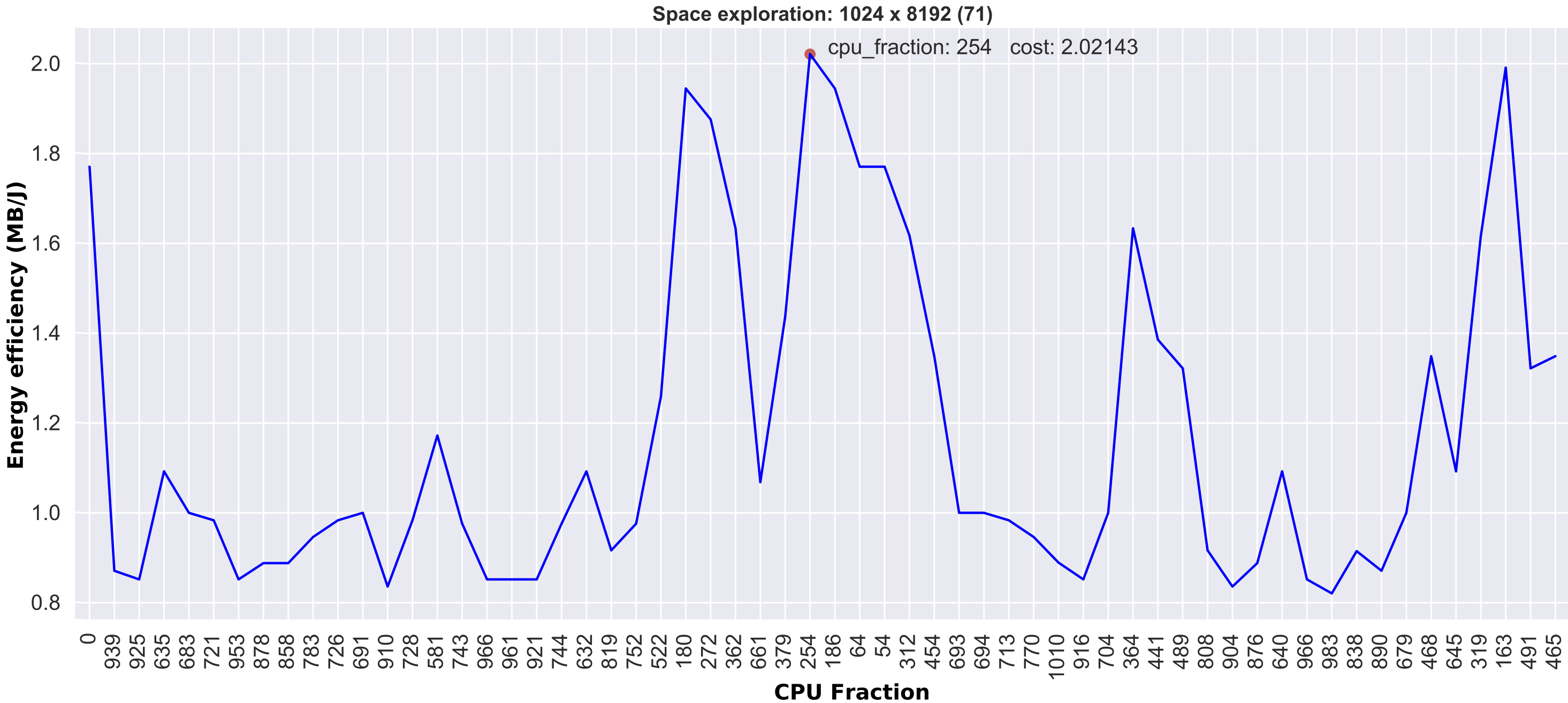}
	\caption{An example of the configuration space exploration for PCC on Ida using our SA-based method. We may observe the presence of multiple local optimums, and that the SA determined the global optimum.}
	\label{fig:space_exploration_example}
\end{figure}

Figure \ref{fig:space_exploration_example} illustrates the process of space exploration using SA for PCC application on Ida system (see Section \ref{sec:pcc}). While there are many local optimums, our SA-based method manages to avoid them and finds the global optimum.  

Our \emph{optimization goal} is to maximize the performance per Joule, which is the rate of computation that can be delivered for every consumed watt. In HPC community, the standard metric for computationally-intensive application is the floating point operations per second (FLOPS). As we target data-intensive applications we use megabytes-per-second (MB/s) as a metric to indicate the application's throughput at run time. Beside the application performance we also consider the energy consumption, therefore the optimization goal metric is indicated by the total amount of data that can be processed for every consumed watt, that is megabytes-per-watt-second (MB/Ws). Section \ref{sec:perf_metrics} introduces evaluation metrics. 

% --------------------------------------------------------------------------
\subsubsection{Using Machine Learning for Evaluation of System Configuration}
\label{sec:ml}

In this section, we describe our supervised machine learning model that is used for evaluation of performance and energy consumption of data-parallel applications. 

Our goal is to build a machine learning model that is able to predict the energy efficiency for a target data-parallel application running on target heterogeneous system based on characteristics of the given workload, and the available resources. Instead of using analytical models that are tightly coupled to a particular problem and environment, we use supervised machine learning that can be trained for various data-parallel programs and heterogeneous systems. 

Preliminary results that compare various machine learning models (including linear regression, Poisson regression, and boosted decision tree regression) have suggested that for this particular type of problem the decision tree regression machine learning algorithm results with highest prediction scores. Hence, the decision tree regression is used in our approach.

We use the python science kit \cite{pedregosa2011scikit} (scikit-learn) to develop our machine learning model. The decision tree regression model is boosted with the Adaboost ensemble method, which results with further improvement with respect to the prediction accuracy. 

To validate the results, we have used two types of well-established validation techniques, the k-fold cross-validation \cite{anguita2012k} (where \textit{k=10}) and the train test split technique (where 80\% of the data is used for training and 20\% of the data is used for testing). In both cases, the model resulted with high (more than 95\%) prediction accuracy scores, which we consider acceptable for the purpose of this study. Note that for evaluating the prediction model, we use the R squared \cite{renaud2010robust} metric, also known as the coefficient of determination, which basically measures how close the data are to the fitted regression line. Also, note that there is still room for improvement with respect to the accuracy of the prediction model, for instance by tuning the hyper parameters, or providing more training data (especially for the data-set for the example of PCC running on GPU), which was out of scope of this paper.

The data pre-processing transforms the string-type input values (such as thread affinity, which value is balanced, scatter, or compact) into integer types (see Table \ref{table:parameter-values} for the full set of parameters and their corresponding values).

%=======================================================================
\section{Evaluation}
\label{sec:evaluation}

In this section, we evaluate experimentally our proposed optimization approach for workload distribution on heterogeneous computing platforms. 

%-------------------------------------------------------------------
\subsection{Evaluation Metrics}
\label{sec:perf_metrics}

In this section, we describe a collection of metrics that we use for performance evaluation. 

\textbf{Time:} is determined by the execution time of the slowest processing unit (that is, CPU or accelerator), 

\begin{equation}
Time (s) = max(CPU_{Time}, ACC_{Time})
\end{equation}

$CPU_{Time}$ is workload processing time on CPU, and $ACC_{Time}$ is workload processing time on accelerator; $ACC_{Time}$ includes also the overhead for transferring data between CPU and accelerator. 

\textbf{Throughput:} is the amount of data processed within a given time,

\begin{equation}
Throughput (MB/s) = Workload\_size/Time
\end{equation}

We can also express the throughput separately for each available processing unit as follows,
\begin{equation}
\begin{aligned}
CPU_{Throughput} (MB/s) = CPU_{Workload\_size} / CPU_{Time} \\
ACC_{Throughput} (MB/s) = ACC_{Workload\_size} / ACC_{Time}
\end{aligned}
\end{equation}

\textbf{Energy:} is the total energy consumed by all used processing unit. We use MeterPU for energy measurement of host CPUs and GPUs, and x-MeterPU \cite{memeti2017benchmarking} for energy measurements on Intel Xeon Phi.

\begin{equation}
Energy (J) = CPU_{Energy} + ACC_{Energy}
\end{equation}

\textbf{Power:} is the consumed energy per time unit. 

\begin{equation}
\begin{aligned}
CPU_{Power} (W) = CPU_{Energy}/CPU_{Time} \\
ACC_{Power} (W) = ACC_{Energy}/ACC_{Time} \\
Power (W) = CPU_{Power} + ACC_{Power} 
\end{aligned}
\end{equation}

$CPU_{Power}$ is power consumed by CPU, $ACC_{Power}$ is power consumed by accelerator, and $Power$ is the total power consumption for workload processing by CPU and accelerator. 

\textbf{Energy Efficiency:} is the ratio of throughput to power,  

\begin{equation}
Energy\_Efficiency (MB/J) = Throughput (MB/s) / Power (W)
\end{equation}

% ------------------------------------------------------------------
\subsection{Performance Comparison of AML and EM}
\label{sec:perf_comparison}

In this section we,

\begin{itemize}
    \item demonstrate experimentally the prediction accuracy of the ML-based performance model for Ida and Emil,
    \item demonstrate that it is significantly faster to evaluate the performance using the ML-based prediction model in comparison to the execution of the real program on Emil and Ida,
    \item demonstrate that AML is capable to determine a near-optimal system configuration with a small fraction of performance experiments in comparison to EM (that is, brute force search). 
\end{itemize}

% ------------------------------------------------------------------
\subsubsection{Prediction Accuracy and Time-efficiency of the ML-based Performance Model}
\label{sec:model-accuracy}

To predict the energy efficiency of PCC application on Ida computing system, we have developed a prediction model based on Boosted Decision Tree Regression. In this section, we address two concerns: (1) are the model-based predicted results comparable to those obtained by measurements, and (2) how much faster is to evaluate the performance using the model compared to measurement?

\begin{figure}[htb]
	\centering
	\includegraphics[width=\linewidth]{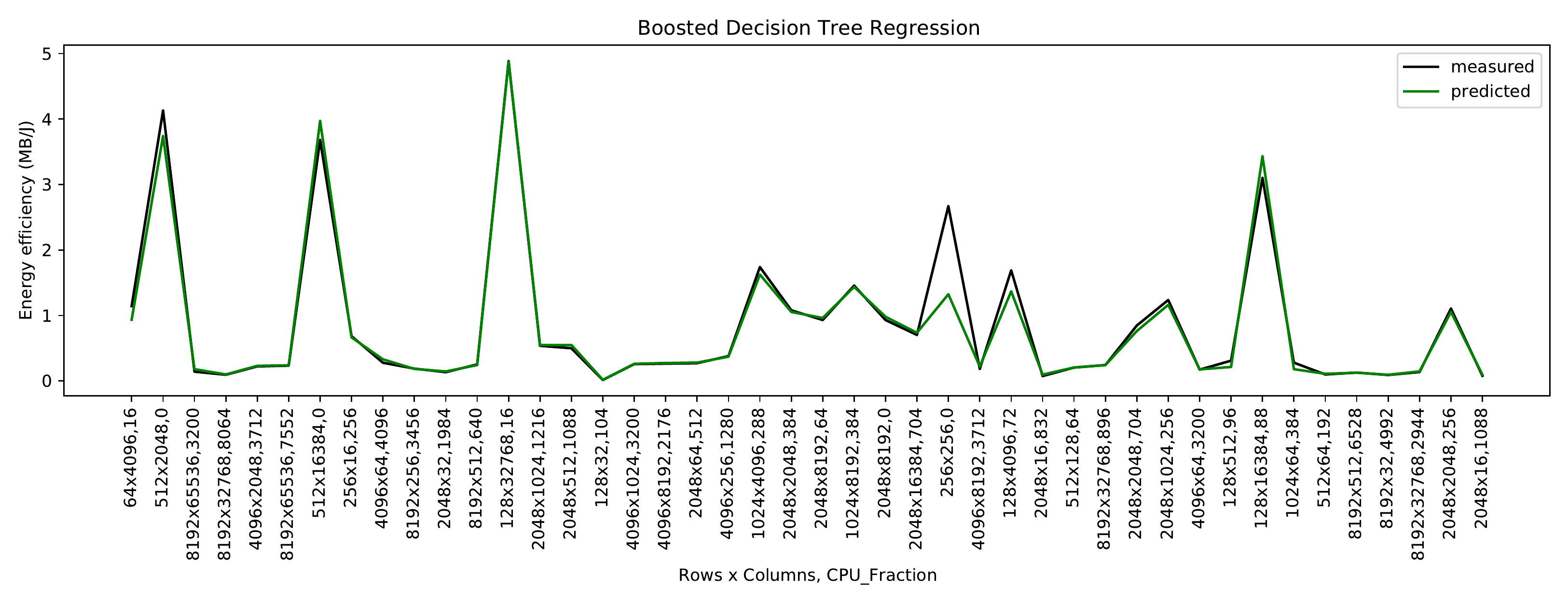}
	\caption{Measured and predicted energy efficiency for PCC application on Ida system. Energy efficiency is predicted for various matrix sizes and fractions of matrix rows processed by CPU using our prediction model built based on Boosted Decision Tree Regression.}
	\label{fig:pred_accuracy}
\end{figure}

Figure \ref{fig:pred_accuracy} depicts measured and predicted energy efficiency for PCC application on Ida system for various matrix sizes and fractions of matrix rows processed by CPU. We may observe that there is good matching of results obtained from the model with the measurement results.

EM involves running PCC on Ida to determine energy efficiency for various parameter values. Execution of 2912 EM experiments in this study took 13404 seconds (approximately 3.7 hours). In contrast to EM, AML uses a machine learning model to predict energy efficiency for various parameter values; model-based evaluation lasts 3.5 milliseconds. Regarding time-efficiency, for 2912 performance experiments EM and AML compare as follows,

\begin{itemize}
    \item EM: 13404 [s]
    \item AML: 10 [s]
\end{itemize}

We may observe that in this study AML is more than 1300 times faster than EM. 

% ------------------------------------------------------------------
\subsubsection{Determining Near-optimal System Configuration for PCC on Ida}
\label{sec:ida-optmimization}

In this section, we compare AML with EM with respect to search for a near-optimal system configuration for PCC application on Ida computing system. The optimization goal is the energy efficiency. Major properties of PCC and Ida are described in Section \ref{sec:sysapps}. AML and EM optimization techniques are described in Section \ref{sec:aml}. 

\begin{figure}[htb]
	\centering
	\includegraphics[width=\linewidth]{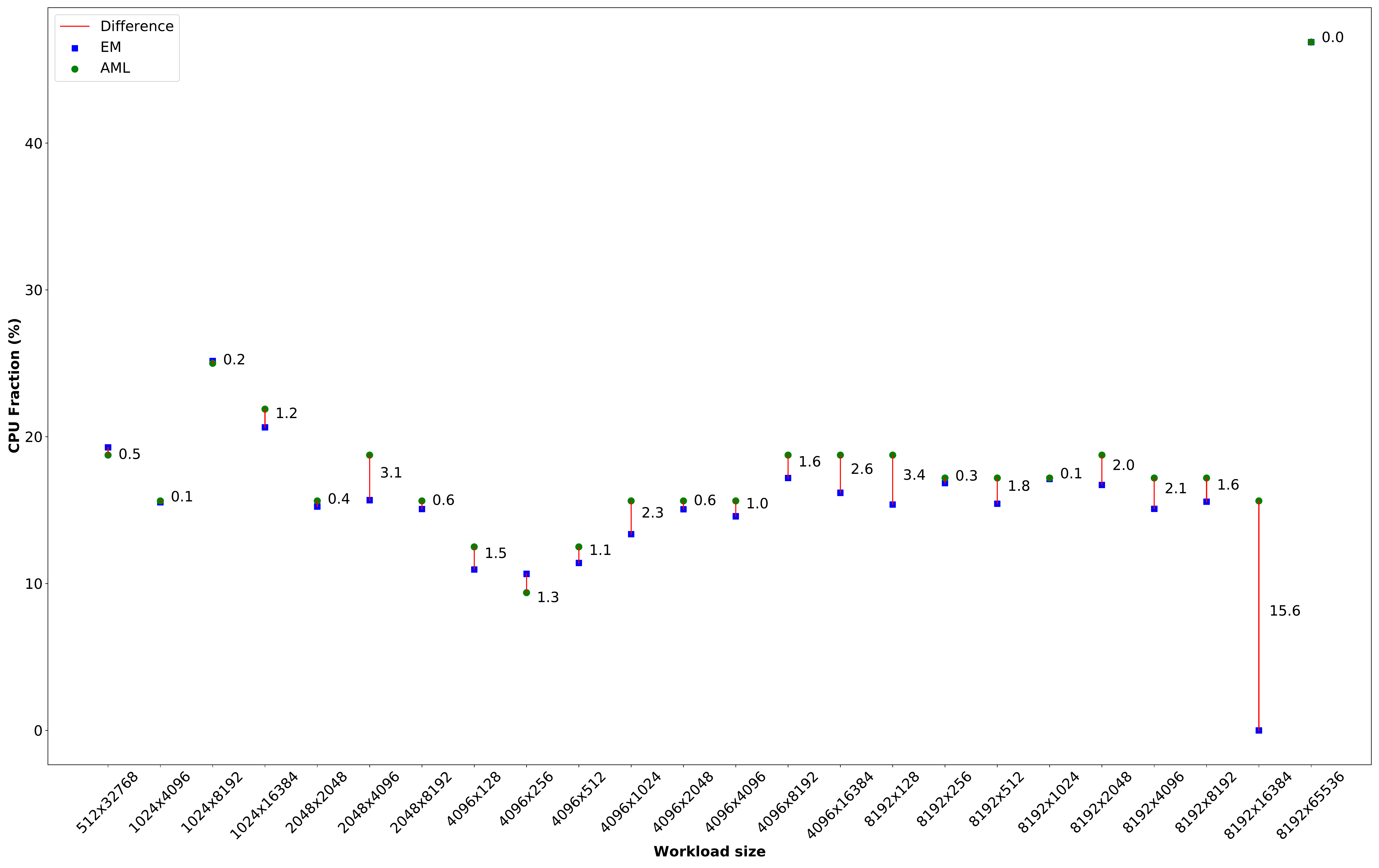}
	\caption{Examples of estimation of near-optimal CPU Fraction with AML and EM for PCC application on Ida. CPU Fraction is the percentage of matrix rows that is mapped for processing to CPU. The optimization goal is the energy efficiency.}
	\label{fig:mml_cpu_fraction}
\end{figure}

We performed experiments for various numbers of matrix rows (16, 32, 64, 128, 256, 512, 1024, 2048, 4096, 8192) and columns (16, 32, 64, 128, 256, 512, 1024, 2048, 4096, 8192, 16384, 32768, 65536). The matrix elements are initialized with random values of float type. Note that the performance results reported in this section do not include the effort required to initialize the matrix. Considering parameter values ranges in this study there are, 

\begin{itemize}
    \item 14677 AML experiments, and
    \item 212914 EM experiments.
\end{itemize}

Figure \ref{fig:mml_cpu_fraction} depicts examples of estimation of near-optimal CPU fraction with AML and EM for PCC application on Ida for various matrix sizes. CPU Fraction is the percentage of matrix rows that is mapped for processing to CPU; the rest of matrix is processed by GPU. The optimization goal is the energy efficiency (see Section \ref{sec:perf_metrics}). 

Please note that the Figure \ref{fig:mml_cpu_fraction} shows only examples of results where the CPU fraction determined by AML differs from the one determined by EM; that is only 24 matrix sizes out of 130 in total. The data points depicted as small rectangles correspond to the EM results, whereas the data points depicted as small circles show the AML values. The vertical red lines show the difference between AML and EM. 

We may observe that in most of the cases AML can determine near-optimal CPU fraction that yields the energy efficiency that is comparable to the one determined by EM. For 106 matrix sizes there was no difference between AML and EM. 

\begin{table}[!t]
	\scriptsize
	\caption{Examples of achieved energy efficiency with EM and AML for various workload sizes of PCC application on Ida.}
	\label{tab:ida_energy_efficiency}
	\centering
  \begin{tabular}{lccc}
	\hline
	Workload: Matrix (R x C) & EM & AML & $\vert Difference\vert$ \\
	\hline
	512 x 32768 & 3.169 & 3.169 & 0.00000 \\
	1024 x 4096 & 2.072 & 2.067 & 0.00474 \\
	1024 x 8192 & 2.021 & 2.021 & 0.00000 \\
	1024 x 16384 & 1.750 & 1.744 & 0.00552 \\
	2048 x 2048 & 1.186 & 1.076 & 0.10936 \\
	2048 x 4096 & 1.079 & 1.059 & 0.01982 \\
	2048 x 8192 & 0.994 & 0.990 & 0.00362 \\
	4096 x 128 & 0.439 & 0.433 & 0.00587 \\
	4096 x 256 & 0.633 & 0.561 & 0.07285 \\
	4096 x 512 & 0.602 & 0.589 & 0.01367 \\
	4096 x 1024 & 0.597 & 0.580 & 0.01656 \\
	4096 x 2048 & 0.569 & 0.547 & 0.02169 \\
	4096 x 4096 & 0.529 & 0.523 & 0.00684 \\
	4096 x 8192 & 0.522 & 0.505 & 0.01669 \\
	4096 x 16384 & 0.487 & 0.487 & 0.00034 \\
	8192 x 128 & 0.233 & 0.211 & 0.02153 \\
	8192 x 256 & 0.252 & 0.243 & 0.00980 \\
	8192 x 512 & 0.280 & 0.268 & 0.01191 \\
	8192 x 1024 & 0.298 & 0.276 & 0.02204 \\
	8192 x 2048 & 0.279 & 0.270 & 0.00907 \\
	8192 x 4096 & 0.268 & 0.260 & 0.00828 \\
	8192 x 8192 & 0.267 & 0.256 & 0.01145 \\
	8192 x 16384 & 0.255 & 0.252 & 0.00338 \\
	8192 x 65536 & 0.529 & 0.529 & 0.00000 \\
	\hline
	\end{tabular}
\end{table}

The highest observed difference is 15.6\% in the case of matrix size 8192x16384; the EM suggests to map all the matrix rows to the GPU (that is, CPU fraction is 0), but the AML suggests to map 15.6\% of the workload to the CPU and the rest to the GPU . While the CPU fraction difference of 15.6\% is substantial, the difference between AML and EM with respect to the achieved optimization goal (that is, energy efficiency) is insignificant:

\begin{itemize}
    \item AML achieved energy efficiency: 0.252 (MB/J)
    \item EM achieved energy efficiency: 0.255 (MB/J)
\end{itemize}

The corresponding energy efficiency results for workload sizes depicted in the Figure \ref{fig:mml_cpu_fraction} are listed in Table \ref{tab:ida_energy_efficiency}. We may observe that the difference between EM and AML with respect to energy efficiency is not significant; that is AML is capable of determining system configurations that result with energy efficiency comparable to EM.

% ------------------------------------------------------------------
\subsubsection{Determining Near-optimal System Configuration for Parallel Pattern Matching on Emil}
\label{sec:emil-optimization}

In this section, we compare AML with EM with respect to search for a near-optimal system configuration for parallel pattern matching on Emil computing system. The optimization goal is the energy efficiency. We have described parallel pattern matching and Emil computing system in Section \ref{sec:sysapps}.

In this study, despite the fact that we tested only what we considered reasonable parameter values (listed on Table \ref{table:parameter-values} in Section \ref{sec:sys}), 5590 experiments were required when we used EM. Our heuristic-guided approach AML that is based on Simulated Annealing and Machine Learning leads to comparatively good performance results, while requiring only a relatively small set of experiments to be performed.

\begin{figure}[!ht]
	\centering
	\includegraphics[width=\linewidth]{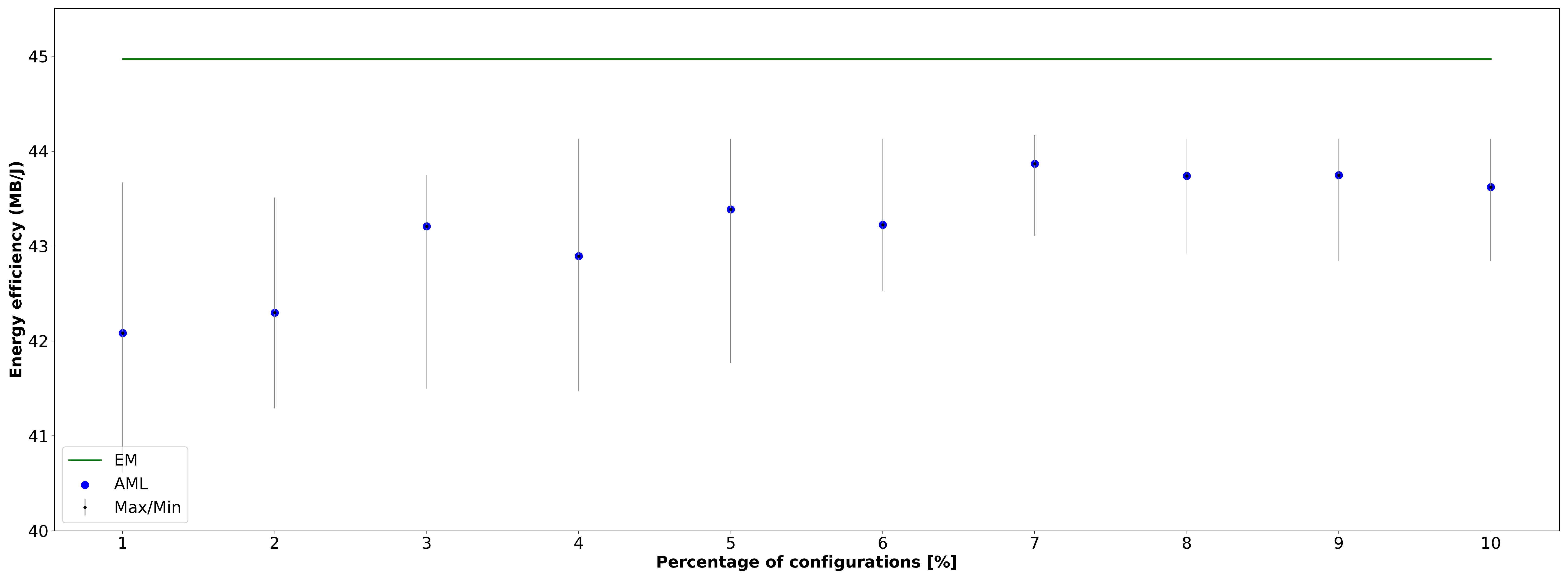}
	\caption{Comparison of AML with EM for various percentages of possible system configurations of Emil. Note that the total number of configurations is 5590 and by trying only about 7\% of all configurations using AML we manage to find a near optimal system configuration that delivers up to 97.55\% of the energy efficiency that is determined by the EM by trying all possible configurations.}
	\label{fig:emil_energy_efficiency}
\end{figure}

Figure \ref{fig:emil_energy_efficiency} depicts the energy efficiency of the pattern matching application when running using the system configuration suggested by AML. The solid horizontal line indicates the energy efficiency of the system configuration determined by EM. EM finds the optimal solution by a kind of brute-force search. We may observe that after evaluating only about 7\% of the possible configurations, AML is able to determine a system configuration that results with energy efficiency that is close to the one of the system configuration determined by EM. The difference between AML and EM with respect to the achieved optimization goal (that is, energy efficiency) is as follows:

\begin{itemize}
    \item AML achieved energy efficiency: 43,87 (MB/J)
    \item EM achieved energy efficiency: 44,97 (MB/J)
\end{itemize}

Please note that AML uses a stochastic search algorithm for global optimization and to avoid ending at a local optima during the search sometimes it accepts a worse system configuration (see Section \ref{sec:SA}). 

%=================================================================
\section{Related Work}
\label{sec:rw}

In this section, first we discuss related research that addresses optimization of heterogeneous computing systems. We are interested to know whether the optimization involves host CPUs and device accelerators (such as, GPU or Intel Xeon Phi), AI heuristics are used for searching for optimal system parameters, machine learning is used for optimization, and whether the optimization goals are power / energy consumption and performance. A summary of related work is provided in Table \ref{table:rw}. Thereafter, we briefly describe how our approach that is presented in this paper differs from the related work.

Huang et al. \cite{huang:2019} propose a strategy for improvement of GPU power consumption based on dynamic voltage and frequency scaling (DVFS). They developed a DVFS model for energy optimization based on proportional-integral-derivative neural network (PIDNN). Three NVIDIA GPUs are used for experimental evaluation: Quadro FX 380, GTX 460, GTX 680. Authors have applied their method only to single GPU systems, and they plan to study multi-GPU systems in future. 

Haidar et al. \cite{haidar2017power} study the relationship between energy consumption and performance for various BLAS kernels on Intel Xeon Phi Knights Landing (KNL). Authors use the PAPI library for power and performance monitoring. A conclusion of this study is that to achieve a good performance with minimal power consumption it is important to maximize the use of MCDRAM memory on KNL device and minimize the use of DDR4 RAM of the system. 

In \cite{kasichayanula2012power}, Kasichayanula et al. propose a software approach for real-time energy consumption measurements and analysis on individual GPU components. They use small subset of real-time statistics, and are able to infer about the activity of other micro-components using a linear-regression based prediction model.

Pereira et al. \cite{Pereira:2020} propose an approach for determining optimal CUDA block dimension for execution of a wind field calculation program on GPU. Authors use particle swarm optimization (PSO) heuristic to find the CUDA block dimension that results with minimal program execution time. Using PSO they achieve 2x speed up compared to the previous GPU version of the program. 

Hong and Kim \cite{hong2010integrated} proposed an integrated energy and performance prediction model for GPU accelerators that is able to predict the optimal number of processors for a given memory-bound application that results with the peak performance and lowest energy consumption. Their experimental results with five various memory bound application benchmarks on a GPU architecture show up to 10\% reduction of energy consumption.

Cerotti et al. \cite{Cerotti:2016} study the performance modeling of GPU accelerated computing systems. Authors present a language for description of  system components and workload elements. Properties of system components are described with a set of parameters. For modeling systems that comprise CPUs and GPUs authors use queuing networks.  

Benkner et al. \cite{benkner11} developed PEPPHER that is a programming framework for heterogeneous systems that comprise CPUs and accelerators (such as, GPU or Intel Xeon Phi). PEPPHER involves source-to-source compilation and a run-time system capable of mapping code components on an extensible set of target processor architectures. To address the performance portability, PEPPHER uses optimized component implementation variants for each execution context. Regression analysis of historical performance data from previous component executions is used to continuously learn associated performance models for run-time scheduling decisions. 

The distribution of computation in heterogeneous systems accelerated with GPU devices for energy optimization has been studied by Ge et al. \cite{peach2014}. Authors propose the so called PEACH model that distributes the computations by splitting the workload between host and accelerators, and adaptively schedules these computations with regards to the computation units speed and energy consumption. Authors use analytical models to optimally determine the workload distribution and scheduling that aims at minimal energy consumption and best performance. 

Ravi and Agrawal \cite{ravi2011dynamic} propose a dynamic scheduling framework for heterogeneous computing systems that comprise various processing elements (such as, CPU or GPU). Authors specifically address data parallel loops. Scheduling decisions consider architectural trade-offs, computation and communication patterns. Performance optimization aims at minimizing data transfer between CPU and GPU, reducing the number of kernel invocation on GPU, and reducing the idle time of resources. 

Grewe and O'Boyle \cite{grewe2011static} study task partitioning for OpenCL programs. Authors address task partitioning and mapping on heterogeneous systems that comprise a CPU and a GPU. Static analysis is used for code features extraction (such as, int and float operations, or data transfer size), and thereafter a machine learning model that is developed based on support vector machines (SVM) is used to map code features to work partition. The model predicts the percentages of work that should be assigned to GPU and CPU to achieve optimal performance.  

\begin{table}[th]
	\centering
	\scriptsize
	\caption{A summary of related work with respect to whether optimization involves host CPUs and device accelerators (such as, GPU or Intel Xeon Phi), heuristic search is used for optimal system parameters, machine learning is used for optimization, and whether the optimization goals are power / energy consumption and performance.}
	\label{table:rw}
	\begin{tabular}{p{2.4cm}p{1cm}p{1cm}p{1.2cm}p{1.2cm}p{1cm}p{1.4cm}}
		\toprule
		Reference 
		& Host & Device & Heuristic search & Machine learning & Power, energy & Performance\\ 
		\midrule
		Huang \cite{huang:2019} 
		& no & yes & no & yes & yes & no\\ 
		Haidar \cite{haidar2017power} 
		& no & yes & no & no & yes & yes\\ 
		Kasichayanula \cite{kasichayanula2012power} 
		& no & yes & no & yes & yes & no\\ 
		Pereira \cite{Pereira:2020}  
		& no & yes & yes & no & no  & yes\\ 
		Hong \cite{hong2010integrated} 
		& no & yes & no & no & yes  & yes\\ 
		Cerotti \cite{Cerotti:2016}  
		& yes & yes & no & no & no  & yes\\ 
		Benkner \cite{benkner11} 
		& yes & yes & no & yes & no  & yes\\ 
		Ge \cite{peach2014} 
		& yes & yes & no & no & yes & yes\\ 
		Ravi \cite{ravi2011dynamic}
		& yes & yes & no & no & no  & yes\\ 
		Grewe \cite{grewe2011static}
		& yes & yes & no & yes & no  & yes\\ 
		This paper 
		& yes & yes & yes & yes & yes & yes \\ 
	\bottomrule
	\end{tabular}
\end{table}

Compared to the related work summarized in Table \ref{table:rw}, in addition to using machine learning for modeling and evaluation of performance and power consumption, we use heuristics to search for a near-optimal system configuration. In a large parameter space for system configuration, heuristics have advantages (such as, time to solution or general applicability) compared to time-prohibitive brute-force approaches that aim at considering all possible configurations or compared to ad hoc approaches that are limited to a specific case.

%===========================================================================
\section{Summary and Future Work}
\label{sec:conclusion}

Heterogeneous architectures are a viable way of building computer systems with a high peak performance and a lower energy consumption. Optimal work-sharing among available CPUs and accelerators is not obvious, and considering both performance and energy consumption further complicates the work-sharing problem. We have presented an approach that uses a probabilistic heuristic search technique for parameter space exploration. The optimization process involves generation of system configuration based on random selection of parameter values and the system performance evaluation using a machine learning model. We have evaluated our approach experimentally on heterogeneous systems that are accelerated with the Intel Xeon Phi or GPU for two applications: Pearson correlation coefficient and parallel pattern matching. In our experiments we have observed that,

\begin{itemize}
    \item the more than 95\% accuracy of the developed machine learning model enables to inform the search for optimal system configuration,
    \item by considering only about 7\% of the possible system configurations our method finds a near-optimal configuration with respect to performance per Joule;
    \item our method is 1300x faster than the brute-force search combined with measurement of program execution;
    \item applicability of our approach is not limited to one type of accelerator or application; the model can be trained for various accelerated systems. 
\end{itemize}

Future work will study work-sharing in the context of the Edge, Fog and Cloud systems.

%===========================================================================
%\bibliographystyle{abbrv} 
%\bibliography{mybib}

\end{document}